\documentclass[journal, letterpaper]{IEEEtran}

\usepackage{graphicx}

\usepackage{url}        

\usepackage{amsmath}

\usepackage{textgreek}	
\usepackage{listings}
\usepackage{csvsimple}
\usepackage{longtable}

\begin{document}

	\title{Automating Hot-Rolling: Designing an Integrated Mechatronics System for Enhanced Efficiency in Sheet Metal Production}
	\author{Mostafa Ahmed, Mohamed Khaled, Abdelrahman Ali, Amr Mostafa, Mariam Mohamed, Omar Ahmed, Osama Khalil}
	\markboth{MTE421 Mechatronics Systems Design}{}
	\maketitle

\section{Introduction}
	
	\PARstart{t}{he} hot-rolling process plays a pivotal role in sheet metal production within heavy steel industries. The conventional approach involves manual adjustment of parameters such as sheet metal velocity and the gap between rotating rolls. However, to enhance efficiency, precision, and automation, our project endeavors to develop an integrated mechatronics system capable of controlling both the speed of the rolling process and the thickness of the sheet metal automatically.

The manual adjustment of parameters in the hot-rolling unit not only introduces inefficiencies but also limits the potential for precise control and optimization. By automating this process, we seek to address these limitations, improving productivity, consistency, and quality in sheet metal production.

This report presents the design and development of the automated hot-rolling mechatronics system, which consists of a pair of rolls applying compression loads to the sheet metal. Our objectives include designing a mechanism for gap control between the rolls, selecting suitable motors and sensors, modeling the system dynamics, simulating its behavior, and proposing practical implementation strategies.

Through this project, we aim to demonstrate the feasibility and effectiveness of integrating mechatronics into the hot-rolling process, offering a comprehensive solution to modernize sheet metal production in heavy steel industries. By providing automated control over critical parameters, such as sheet metal thickness and production speed, our system promises to optimize performance, enhance product quality, and streamline manufacturing operations.

The following sections will detail our design approach, component selection rationale, mathematical modeling, simulation results, and practical implementation considerations. This project represents a significant step towards realizing the potential of mechatronics in industrial automation, with implications for improving efficiency and competitiveness in the sheet metal production sector.

\section{Explanation of the Mechanism System Design}

In the quest to enhance the efficiency and precision of the hot-rolling process, we propose the integration of a power screw mechanism to automate the control of the gap between the rolls. This innovative approach offers several advantages over traditional manual adjustment methods, including increased accuracy, repeatability, and ease of control.

\textbf{Idea Discussion:}

The power screw mechanism operates on the principle of converting rotary motion into linear motion through the interaction of a threaded shaft (the screw) and a nut. By rotating the screw, the nut moves along its threads, thereby translating linearly. This linear motion can be utilized to precisely adjust the gap between the rolls in the hot-rolling unit.

One of the key advantages of employing a power screw mechanism is its ability to provide fine-grained control over the gap distance. By selecting an appropriate pitch for the screw threads and coupling it with a high-resolution rotary actuator, we can achieve sub-millimeter adjustments with ease. This level of precision is essential for meeting stringent quality standards and accommodating varying sheet metal thicknesses.

Moreover, the power screw mechanism offers inherent mechanical advantage, allowing for the generation of substantial axial force with relatively low input torque. This ensures that the gap adjustment can be performed efficiently even under heavy loads, such as when compressing thick sheet metal.

\begin{figure}[h!]
    \centering
    \includegraphics[scale=0.5]{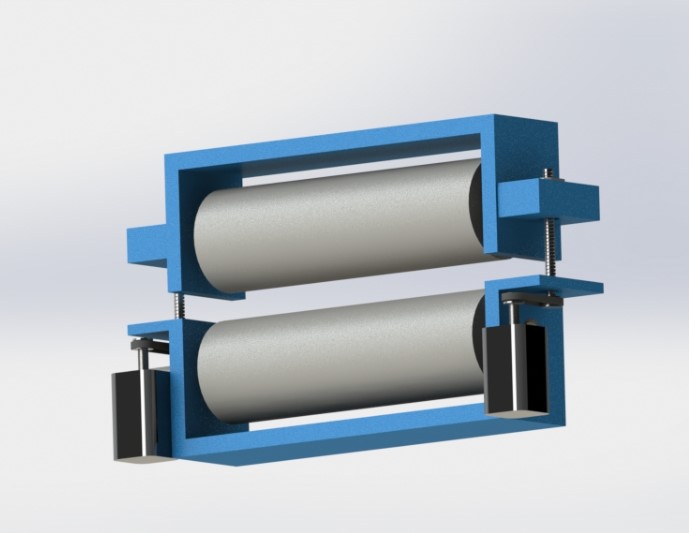}
    \caption{Our Design from SolidWorks}
\end{figure}

\pagebreak

\textbf{Design Considerations:}

In designing the power screw mechanism, several factors must be taken into account to ensure optimal performance and reliability:

\begin{enumerate}
    
    \item \textbf{Screw Pitch and Lead:} The selection of an appropriate screw pitch determines the linear distance traveled per revolution of the screw. A finer pitch allows for finer adjustments but may require higher torque to actuate.
    
    \item \textbf{Actuation System:} The rotary actuator responsible for driving the screw must be carefully chosen based on torque requirements, speed, and precision. Stepper motors or servo motors with high-resolution encoders are often suitable for this application.
    
    \item \textbf{Nut Design:} The design of the nut that engages with the screw threads plays a crucial role in ensuring smooth and stable motion. Anti-backlash nuts or preloaded ball screws may be employed to minimize play and ensure repeatability.
    
    \item \textbf{Support and Guidance:} Proper support and guidance mechanisms are necessary to maintain the alignment and stability of the screw shaft. Linear bearings or guide rails may be utilized to minimize deflection and ensure smooth operation.
    
    \item \textbf{Safety and Reliability:} Measures should be implemented to prevent overloading of the mechanism and to ensure safe operation under all conditions. Limit switches, torque sensors, and emergency stop mechanisms may be incorporated for added safety.

\end{enumerate}

By integrating a power screw mechanism for gap control, we aim to revolutionize the hot-rolling process, enhancing its efficiency, precision, and automation capabilities. This innovative solution represents a significant advancement in mechatronics-driven industrial automation, with the potential to revolutionize sheet metal production in heavy steel industries.

\section{Force Analysis and Motor Selection}

We must determine the necessary torque and speed (RPM) of the motor based on the force analysis to choose the right motor for the roll's primary rotation motion in the hot-rolling unit. The procedures and formulas required for these computations are listed below:

\subsection{Force Analysis}

\subsubsection{Compression Force Calculation}

Firstly, we calculate the compression force \( F \) required to reduce the thickness of the sheet metal. The force can be determined using the yield strength \( \sigma_y \) and the contact area \( A \):

\[ F = \sigma_y \times A \]

Given:
\begin{itemize}
    \item Yield strength of hot-rolled steel \( \sigma_y = 150 \, \text{MPa} = 150 \times 10^6 \, \text{N/m}^2 \)
    \item Sheet width \( w = 1 \, \text{m} \)
    \item Initial thickness \( t_i = 5 \, \text{mm} = 0.005 \, \text{m} \)
    \item Final thickness \( t_f = 1 \, \text{mm} = 0.001 \, \text{m} \)
    \item Roll diameter \( D \) between 200mm and 300mm (Assume \( D = 0.25 \, \text{m} \))
\end{itemize}

The contact length \( L \) between the roll and the sheet metal can be approximated assuming the deformation region forms a rectangle as follows:

\[ L = D \times \arcsin \left( \frac{t_i - t_f}{D} \right) \approx D \times \frac{t_i - t_f}{D} = t_i - t_f \]

Thus, the contact area \( A \) is:

\[ A = w \times L = 1 \times (t_i - t_f) = 1 \times (0.005 - 0.001) = 0.004 \, \text{m}^2 \]

The compression force \( F \) is:

\[ F = 150 \times 10^6 \times 0.004 = 600,000 \, \text{N} \]

\subsection{Torque Calculation}

Considering the force applied at the contact region and the rolls' radius, the torque \( T \) needed to spin the rolls may be computed as:

\[ T = F \times \frac{D}{2} \]

\[ T = 600,000 \times 0.125 = 75,000 \, \text{Nm} \]

\subsection{Motor Power Calculation}

The power \( P \) required by the motor is given by:

\[ P = T \times \omega \]

Where \( \omega \) is the angular velocity in rad/s. The angular velocity is related to the linear velocity \( v \) of the sheet metal and the radius of the rolls as follows:

\[ v = r \times \omega \]

\[ \omega = \frac{v}{r} = \frac{0.5 \, \text{m/s}}{0.125 \, \text{m}} = 4 \, \text{rad/s} \]

Thus, the power \( P \) is:

\[ P = 75,000 \times 4 = 300,000 \, \text{W} = 300 \, \text{kW} \]

\subsection{Gear Reduction}

A gear reduction is frequently required to match the motor speed to the required roll speed. The base speed of the motor may be much faster than the required speed for the roll. One way to compute the speed reduction ratio \( R \) is:

\[ R = \frac{\text{Motor Speed}}{\text{Roll Speed}} \]

Assume a motor speed of 1500 RPM and required roll speed calculated from \( \omega \):

\[ \text{Roll Speed} = \frac{4 \, \text{rad/s}}{2\pi} \times 60 \approx 38 \, \text{RPM} \]

Thus, the gear reduction ratio \( R \) is:

\[ R = \frac{1500}{38} \approx 39.5 \]

\subsection{Gear Head Selection}

A reduction ratio of roughly 40:1 for a gear head will work well to get the needed roll speed out of the motor speed.

\section{Components Choices and Justification}

\subsection{Motor Selection}

We need to select a motor with a minimum power output of 300 kW. For such high-power requirements, industrial motors such as DC motors or AC induction motors are appropriate. While AC motors usually offer greater energy economy and are easier to maintain, DC motors are distinguished by their high starting torque and precise speed control.

\subsubsection{Type of the selected Motor: High-Power AC Induction Motor}

\paragraph{Justification:}

\begin{itemize}
    \item \textbf{High Efficiency and Reliability}: In industrial applications, AC induction motors are known for their exceptional performance and reliability. They are strong and able to withstand the harsh conditions of a hot-rolling mill where high-power continuous operations are necessary.
    \item \textbf{High Power Capability}: It is simple to develop AC induction motors to provide the high power needed. A motor with a power rating of at least 300 kW is required for this application. It is possible to scale AC induction motors effectively to satisfy these power needs.
    \item \textbf{Ease of Maintenance}: These motors are relatively easy to maintain compared to other high-power motor types, which is crucial for minimizing downtime in industrial settings.
    \item \textbf{Cost-Effectiveness}: AC induction motors are generally more cost-effective for high-power applications compared to equivalent DC motors, both in terms of initial cost and operational cost.
\end{itemize}

\paragraph{Specifications:}
\begin{itemize}
    \item Power: 300 kW
    \item Speed: 1500 RPM
    \item Voltage: Typically 400-690V (three-phase)
    \item Efficiency: >90%
\end{itemize}

For these specifications, we suggest the W22 IE4 300 kW 4P 355M/L 3Ph 400/690//460 V 50 Hz IC411 - TEFC - B14T.

\subsection{Gear Head Selection}

\subsubsection{Type of the Gear Head selected: Industrial Gearbox with Reduction Ratio 40:1}

\paragraph{Justification:}

\begin{itemize}
    \item \textbf{Speed Reduction}: The rolls require a speed of about 38 RPM, whereas the motor runs at 1500 RPM. A 40:1 gear reduction ratio efficiently matches the motor speed to the necessary roll speed.
    \item \textbf{High Torque Handling}: Gearboxes are designed to handle high torque loads, which is essential for transmitting the required torque from the motor to the rolls.
    \item \textbf{Durability}: The extreme temperatures found in a hot-rolling mill are only one of the tough circumstances that industrial gearboxes are designed to handle.
\end{itemize}

\paragraph{Specifications:}
\begin{itemize}
    \item Reduction Ratio: 40:1
    \item Input Speed: 1500 RPM
    \item Output Speed: 37.5 RPM (approximately 38 RPM)
    \item Output Torque: The torque generated by the motor must be handled by the gearbox and amplified using the reduction ratio.
\end{itemize}

For these specifications, we suggest the Worm gear unit size 040 ratio 40:1 with 63B14 flange.

\subsection{Motor Drive Selection}

\subsubsection{Type of Motor Drive: Variable Frequency Drive (VFD)}

\paragraph{Justification:}

\begin{itemize}
    \item \textbf{Speed Control}: VFDs allow precise control of the motor speed, which is essential for maintaining the desired speed of the rolls and adjusting the speed as necessary for different thicknesses of the sheet metal.
    \item \textbf{Energy Efficiency}: VFDs improve energy efficiency by matching the motor speed to the load requirements, reducing energy consumption during lower load periods.
    \item \textbf{Soft Start and Stop}: VFDs provide soft starting and stopping capabilities, which reduce mechanical stress on the motor and the gear system, enhancing the lifespan of the equipment.
\end{itemize}

\paragraph{Specifications:}
\begin{itemize}
    \item Power Rating: At least 300 kW to match the motor.
    \item Input Voltage: Compatible with the motor (e.g., 400–690V three-phase).
    \item Control Method: Vector control or direct torque control for precise speed and torque management.
\end{itemize}

\paragraph{Equations and Calculations:}

Speed Control:
\[ \text{Speed} = \frac{120 \times \text{Frequency}}{\text{Number of Poles}} \]

For a 4-pole motor:
\[ \text{Speed} = \frac{120 \times \text{Frequency}}{4} \]

To achieve 1500 RPM (25 Hz for 4-pole motor):
\[ \text{Frequency} = \frac{1500 \times 4}{120} = 50 \, \text{Hz} \]

Suggested Drive: Parker SSD 890SD 250kW/300kW 400V Inverter Drive STO C3 EMC DBR 115V Fan.

\pagebreak

\section{Modeling}

To fully model the system we need to control two factors:
\begin{itemize}
    \item Speed of the rolling (Feed)
    \item Thickness of the sheet
\end{itemize}

\subsection{Speed Control}

\subsubsection{Motor Dynamics for Roller Speed}

The speed of the rollers is controlled by a motor. The motor’s transfer function can be modeled as follows:
\[
\omega(s) = \frac{K}{Js + B} \cdot V(s)
\]
Where:
\begin{align*}
\omega(s) & = \text{Angular velocity of the motor} \\
K & = \text{Motor constant} \\
J & = \text{Moment of inertia} \\
B & = \text{Damping coefficient} \\
V(s) & = \text{Input voltage to the motor}
\end{align*}

\subsubsection{Motor Dynamics for Linear Sheet Speed}

The angular velocity is transferred into linear velocity in the input sheet metal as follows:
\[
v(s) = r \cdot \omega(s)
\]
where \( r \) is the radius of the roller.

\subsubsection{Open Loop Transfer Function}

By combining both equations we get the Open Loop Transfer Function:
\[
v(s) = r \cdot \frac{K}{Js + B} \cdot V(s)
\]
where \( v(s) \) is the linear velocity of the input sheet.

The full speed-control Controller Block Diagram:

\begin{center}
    \begin{figure}[h!]
        \centering
        \includegraphics[scale=0.5]{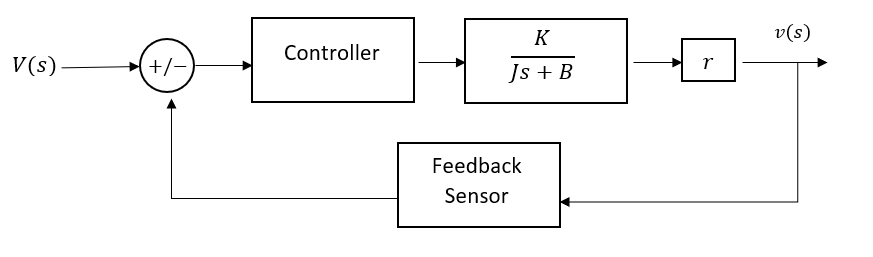}
    \end{figure}
\end{center}
 
\subsection{Modeling the Thickness Control}

The proposed system is simplified by not taking into consideration:
\begin{itemize}
    \item Material Deformation
    \item Spring-Back Effect
    \item Temperature and Material Properties
\end{itemize}
Thus, the final output thickness is directly determined by the gap set by the power screw mechanism.

\subsubsection{Power Screw Dynamics for Gap Control}

\[
x(s) = \frac{L \cdot \omega_{ps}(s)}{2\pi}
\]
Where:
\begin{align*}
x(s) & = \text{Linear displacement (gap between rollers/thickness)} \\
L & = \text{Lead of the screw} \\
\omega_{ps}(s) & = \text{Angular velocity of the power screw}
\end{align*}

\subsubsection{Motor Dynamics for Power Screw}

The motor driving the power screw has a similar transfer function:
\[
\omega_{ps}(s) = \frac{K_{ps}}{J_{ps}s + B_{ps}} \cdot V_{ps}(s)
\]
Where:
\begin{align*}
K_{ps} & = \text{Power screw motor constant} \\
J_{ps} & = \text{Moment of inertia of the power screw motor} \\
B_{ps} & = \text{Damping coefficient of the power screw motor} \\
V_{ps}(s) & = \text{Input voltage to the power screw motor}
\end{align*}

\subsubsection{Gap (Thickness) Control}

Combining the motor and power screw dynamics we get the Transfer Function:
\[
x(s) = \frac{L \cdot K_{ps}}{2\pi \cdot (J_{ps}s + B_{ps})} \cdot V_{ps}(s)
\]

The full gap/thickness-control Controller Block Diagram:

\begin{center}
    \begin{figure}[h!]
        \centering
        \includegraphics[scale=0.5]{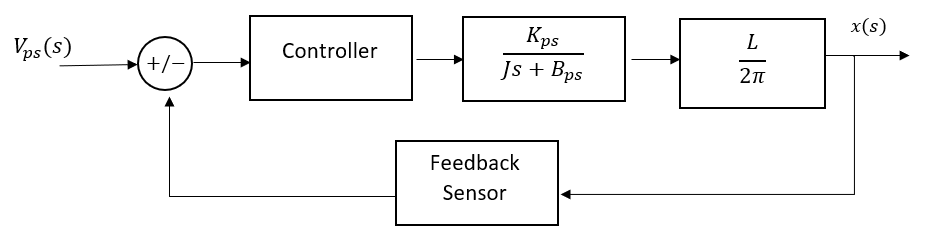}
    \end{figure}
\end{center}

\vspace{-25pt}

\section{Electronics System Design}

In our automated system, electronics play a pivotal role in governing the power screw mechanism, overseeing sensor data, and executing feedback control algorithms. To ensure seamless real-time control and efficient data processing, the selection of electronics components, including microcontrollers or programmable logic controllers (PLCs), is paramount.

A cornerstone in our electronics component selection is the implementation of the PID control algorithm, a fundamental tool for closed-loop control systems:
\[
u(t) = K_p e(t) + K_i \int_{0}^{t} e(\tau) d\tau + K_d \frac{de(t)}{dt}
\]

This equation underscores the significance of choosing microcontrollers or PLCs equipped with robust computational power and ample I/O capabilities. It is imperative to select microcontrollers or PLCs that can effectively execute and fine-tune the PID control algorithm to ensure precise control over the power screw mechanism.
\vspace{5pt}

For instance, opting for PLCs such as the Siemens SIMATIC S7-1500 series or the Allen-Bradley ControlLogix series provides the computational muscle and flexibility required for sophisticated control tasks. These PLCs offer a plethora of digital and analog I/O channels, along with advanced programming capabilities, making them ideal candidates for our automation system's control unit. Similarly, microcontrollers like the Arduino Due or the Raspberry Pi 4, with their high-speed processors and ample memory, can serve as cost-effective alternatives for less complex applications, albeit with slightly lower performance.

\section{Multi-body Dynamics Simulation}
\[Transfer Function = \frac{1}{s^8 + 3.571s^6 + 1}\]

\begin{center}
    \textbf{Before PID}
    \includegraphics[scale=0.35]{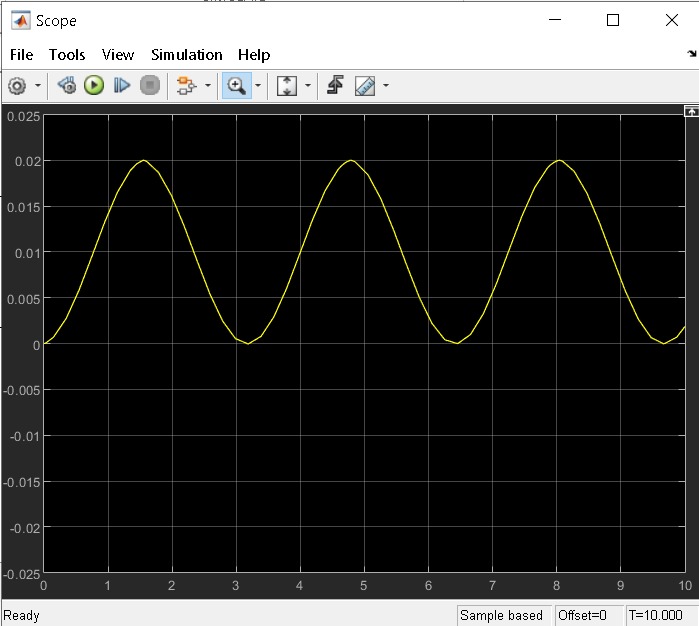}
\end{center}

\begin{center}
    \textbf{After PID: Kp = 0.00941, Ki = 6.53e-05, Kd = 0.339}
    \includegraphics[scale=0.35]{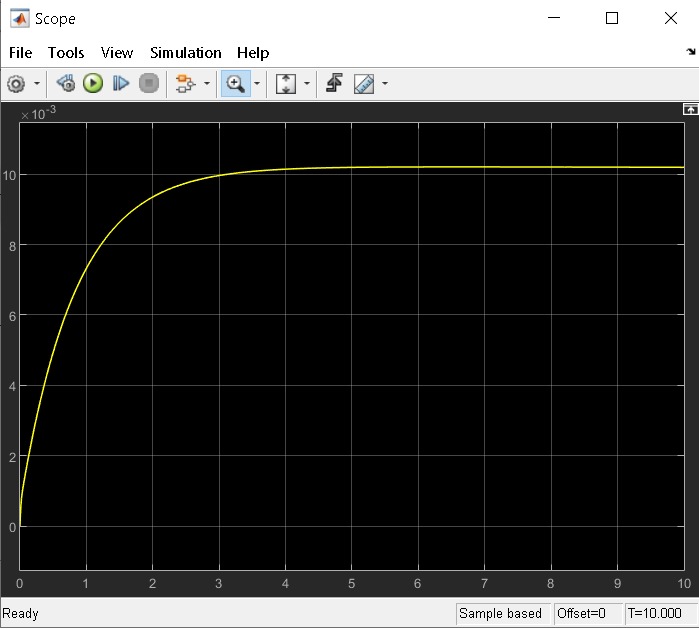}
\end{center}

\begin{center}
    \textbf{Multi-body model}
    \includegraphics[scale=0.25]{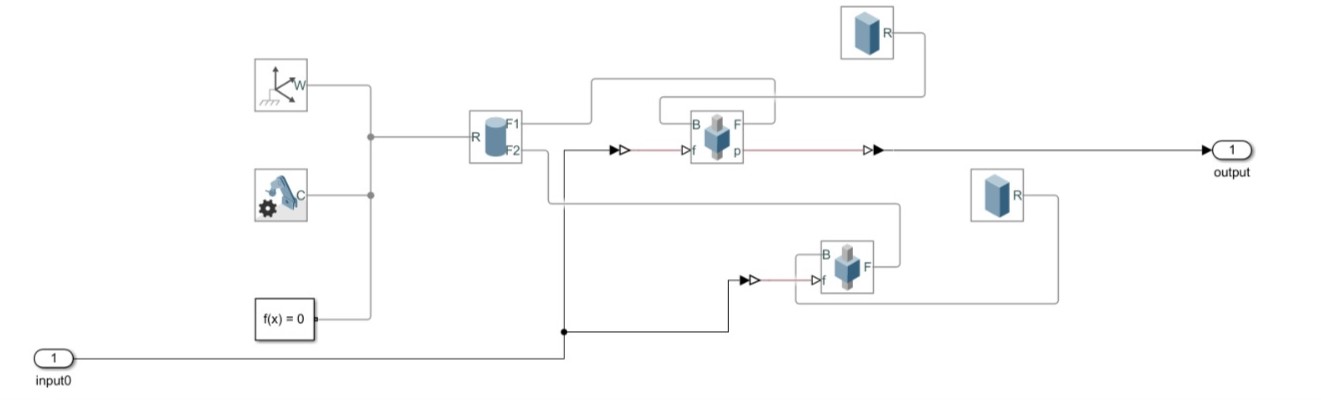}
\end{center}

\pagebreak

\section{Implementing Fault Detection Procedures}
Implementing fault detection procedures in our automated hot-rolling system is crucial for ensuring its safe and reliable operation. Our strategy involves proactive monitoring, real-time diagnostics, and contingency planning to mitigate potential faults effectively. This comprehensive approach begins with the integration of sensors throughout the system, including temperature sensors for monitoring sheet metal and roller temperatures, torque sensors for measuring applied torque to the rolls, and position sensors like linear variable differential transformers (LVDTs) or optical encoders for tracking roll and sheet metal displacement.
\vspace{5pt}

Key to our fault detection strategy is the integration of laser displacement sensors for monitoring sheet metal thickness. By accurately measuring the distance between the sensor and the sheet metal surface, these sensors provide real-time thickness data, crucial for maintaining product quality and process control. This data is fed back to the control unit, enabling dynamic adjustment of the roll gap to ensure consistent thickness reduction. Additionally, the precise position feedback provided by LVDTs or optical encoders allows for real-time monitoring of roll alignment and sheet metal displacement, further enhancing process control and fault detection capabilities.
\vspace{5pt}

By implementing these sensor types and fault detection procedures, we enhance the reliability and productivity of our automated hot-rolling system. Real-time monitoring and dynamic control enable us to detect and address potential faults promptly, minimizing downtime and ensuring consistent product quality. This proactive approach to fault detection is essential for maintaining operational efficiency and safeguarding against disruptions in production.

\section{Demonstrating Feasibility and Practicality}
In evaluating the feasibility and practicality of our design compared to traditional methods, it's essential to highlight the tangible benefits our automated system offers over manual processes.
\vspace{5pt}

Firstly, the manual adjustment of gap settings between rolls in traditional systems is inherently time-consuming. Operators must manually turn handwheels or levers to achieve the desired gap, a process that can be laborious and inefficient. In contrast, our automated system streamlines this process, allowing for seamless, precise adjustments at the push of a button or through programmed commands. This significant reduction in manual labor translates directly into time savings and increased operational efficiency.
\vspace{5pt}

Secondly, precision is paramount in hot-rolling processes to ensure consistent product quality and adherence to specifications. Manual adjustments, reliant on operator skill and judgment, often result in variations in gap settings and, consequently, sheet metal thickness. Our automated system, driven by precise actuators and controlled by sophisticated algorithms, ensures consistent and precise gap adjustments, leading to tighter tolerances and improved product quality.
\vspace{5pt}

Furthermore, the reliability of manual systems can be compromised by human error, fatigue, and inconsistency. Operators may inadvertently deviate from prescribed procedures or fail to maintain consistent operating conditions, leading to inconsistencies in product quality and increased downtime due to maintenance and troubleshooting. In contrast, our automated system operates reliably and consistently under diverse operating conditions, minimizing the risk of errors and ensuring uninterrupted production.

\section{Refrences}
\begin{enumerate}
    \item The provided file: Mini-Project/ Design Mechatronics System of Controlling Hot-Rolling Process of Sheet Metals
    \item \url{https://northendelectric.com/2024/02/14/which-motor-is-more-powerful-ac-or-dc-motors/#:~:text=While%20AC%20motors%20generally%20offer%20greater%20energy%20efficiency,are%20important%20metrics%20that%20should%20guide%20your%20choice.}
    \item \url{https://www.weg.net/catalog/weg/FI/en/Electric-Motors/Low-Voltage-IEC-Motors/Three-Phase/W22-/W22-IE4-300-kW-4P-355M-L-3Ph-400-690-460-V-50-Hz-IC411---TEFC---B14T/p/13004563}
    \item \url{https://en.challengept.com/worm-gear-unit-size-040-ratio-40-1-with-63b14-flange.html}
    \item \url{https://www.advancedenergy.org/news/variable-frequency-drives-and-their-importance-in-industry}
    \item \url{https://inverterdrive.com/group/AC-Inverter-Drives-400V/Parker-SSD-890SD-4-0480H-B-1F-A-UK-00-250kW-300kW/}
\end{enumerate}
\end{document}